\documentclass[prl, twocolumn, amsmath,amssymb, superscriptaddress, nofootinbib, floatfix ]{revtex4-1}

\usepackage{multirow}
\usepackage{graphicx}
\usepackage{amsfonts,tikz}  
\usetikzlibrary{matrix}
\usepackage{epstopdf}
\usepackage[version=3]{mhchem}
\usepackage{color}
\usepackage{hyperref}
\usepackage{enumitem}
\usepackage[scr=boondoxo,scrscaled=1.25]{mathalfa}
\usepackage{scalerel}

\usepackage{ulem} 

\hypersetup{%
  pdfpagemode=None, 
  pdfstartpage=1,
  pdfstartview=FitH,
  pdfmenubar=true,
  pdftoolbar=true,
  colorlinks = true,
  linkcolor=blue,
  citecolor=blue,
  urlcolor = blue,
  bookmarksopen=false
}

\newcommand{\op}[1]{\ensuremath{\mathsf{\hat{#1}}}}

\newcommand{\Bra}[1]{\ensuremath{\left\langle #1 \right\vert}}
\newcommand{\KetBra}[2]{\Ket{#1}\kern-0.1em\Bra{#2}}
\newcommand{\Ket}[1]{\ensuremath{\left\vert #1 \right\rangle}}

\definecolor{Pantone268}{cmyk}{0.82,1.0,0.0,0.12}
\definecolor{HunterOrange}{cmyk}{0.0,0.55,1.0,0.0}
\definecolor{lblue}{cmyk}{1.0,0.13,0.0,0.44}
\definecolor{FUgreen}{cmyk}{0.185,0.0,0.91,0.0}
\definecolor{FUblue}{cmyk}{1.0,0.72,0.0,0.185}

\begin{document}


\title{
Continuum-electron interferometry for enhancement of photoelectron
circular dichroism and measurement of bound, free, and mixed contributions to
chiral response}     

\date{\today}

\author{R.~Esteban Goetz}
\altaffiliation{Present Address: Department of Physics, University of Connecticut, Theoretical Atomic, Molecular and Optical Physics Center, Connecticut, 06269, USA }
\affiliation{J. R. Macdonald Laboratory, Department of Physics, Kansas State University, Manhattan, KS 66506 }

\author{Alexander Blech}
\affiliation{Dahlem Center for Complex Quantum Systems and Fachbereich Physik, Freie Universit\"at Berlin, Arnimallee 14, D-14195 Berlin, Germany}


\author{Corbin Allison}
\affiliation{J. R. Macdonald Laboratory, Department of Physics, Kansas State University, Manhattan, KS 66506 }

\author{Christiane P. Koch}
\affiliation{Dahlem Center for Complex Quantum Systems and Fachbereich Physik, Freie Universit\"at Berlin, Arnimallee 14, D-14195 Berlin, Germany}

\author{Loren Greenman}
\email{lgreenman@phys.ksu.edu}
\affiliation{J. R. Macdonald Laboratory, Department of Physics, Kansas State University, Manhattan, KS 66506 }

\begin{abstract}
  We develop photoelectron interferometry based on laser-assisted extreme 
  ultraviolet ionization for flexible and robust control of photoelectron circular dichroism in randomly oriented chiral molecules. 
A comb of XUV photons ionizes a sample of chiral molecules in the presence of a time-delayed infrared or visible laser pulse promoting interferences between 
components of the XUV-ionized photoelectron wave packet. 
In striking contrast to multicolor phase control schemes relying on pulse shaping techniques, the magnitude of the resulting chiral signal is here controlled by the time delay between the XUV and laser pulses. Furthermore, we show that the relative polarization configurations of the XUV and IR fields allows for disentangling the contributions of bound and
continuum states to the chiral response. Our proposal provides a simple, robust and versatile tool for the control of photoelectron circular dichroism and experimentally feasible protocol for probing the individual contributions of bound and continuum states to the PECD in a time-resolved manner.
\end{abstract}


\maketitle

Chirality, the property of a molecule to be non-superposable with its mirror 
image~\cite{mcnaught1997compendium}, is a ubiquitous property in nature with profound effects on many fundamental chemical and physical processes.
Photoelectron circular dichroism (PECD) is a signature of chirality and designates the relatively strong effect 
whereby randomly oriented chiral molecules ionize asymmetrically for left- and right-circularly polarized light~\cite{RitchiePRA76}. The strength of PECD has led to its use to observe chirality in
molecules~\cite{LuxAngewandte12,LuxCPC15,LuxJPB16,KastnerChemPhysChem16,KastnerJCP17,LehmannJCP13,JanssenPCCP14,
FanoodNatureComm15,FanoodPCCP15,CireasaNatPhys15,CombyJPCL16,BeaulieuFaraday16,BeaulieuNJP16,BeaulieuSci17,BeaulieuNatPhys18,Miles2017,BoeweringPRL01,HergenhahnJCP2004}
and turned it into a practical tool for, e.g., measuring enantiomeric 
excess~\cite{KastnerChemPhysChem16,BeaulieuNatPhys18,Miles2017}.
Techniques to enhance the PECD in chiral molecules  
include time-resolved pump-probe 
photoexcitation circular dichroism~\cite{BeaulieuNatPhys18}, coherent control of PECD 
in bichromatic~\cite{DemekhinPRL2018,RozenPRX2019} and resonantly-enhanced multiphoton ionization~\cite{GoetzPRL2019}, as well as use of a continuum resonance and Fano interference~\cite{HartmannPRL2019}.  
While these techniques focus on maximizing the PECD signal, the relative contribution of bound and continuum states to the PECD has not yet been addressed, and none of these approaches 
have been shown to disentangle the contributions 
of bound and continuum states in a systematic way.
Furthermore, schemes that show high enhancements of the PECD such as the
shaped resonantly-enhanced multiphoton ionization (REMPI) pulses we have
previously suggested~\cite{GoetzPRL2019} require challenging
generation of and control over the spectral coherence of multicolor ultraviolet laser fields.

In this letter, we propose a flexible, yet robust, photoelectron interferometry approach to probe bound and continuum contributions to PECD in randomly oriented molecules.  Our method has
the advantage over that of Ref.~\cite{GoetzPRL2019} that it does not require shaping circularly polarized XUV laser 
sources, a formidable experimental challenge. Rather, it
relies on an adaptation of the RABBITT (reconstruction of attosecond beating by interference of two-photon 
transitions) technique~\cite{paul2001observation,muller2002reconstruction}, a versatile interferometric approach which has been implemented in many experimental
laboratories~\cite{mazza2014determining,donsaPRL2019,klunder2011probing}. 
The necessary circularly polarized XUV fields have been synthesized~\cite{HanNatPhys22,HanOptica23} and used in PECD experiments recently~\cite{WatersCPC22}. A variant of {RABBITT}\- called CHABBIT, a laser-assisted, self-referenced, above-threshold ionization attosecond photoelectron interferometry has been used to extract the temporal profile and phase difference in the photoelectron 
emission in chiral molecules along different directions~\cite{BeaulieuSci17}.
Here, we suggest to directly probe the individual contributions of bound and continuum states
 and 
 their interference to the PECD. We show how 
changing the polarization configurations of the XUV and IR fields results in different asymmetries in the photoelectron angular distributions in randomly oriented chiral molecules.

Using  $\ce{CHBrClF}$ as a prototypical chiral molecule~\cite{PitzerScience13,FehrePRL2021}, we explain how promoting interferences in the continuum in a controlled manner results in precise control of  PECD after XUV ionization in the presence of a time-delayed IR field.
Rather than relying on multiple resonantly enhanced multiphoton ionization and multicolor pulse shaping techniques to control PECD~\cite{GoetzPRL2019}, control here is achieved by adjusting the time delay between the XUV comb and IR field or by varying the IR pulse duration. Changing the polarization configuration of the pulses with optimized delay and IR duration allows for revealing the chiral contributions of different states. 

The photoionization dynamics is described within the fixed-nuclei and nonrelativistic dipole approximations. The time-dependent Schr\"odinger equation reads
\begin{subequations}
\begin{eqnarray}
\label{eq:TDSE}
  i\dfrac{\partial}{\partial\,t}|\Psi^{N}_{\gamma_{\mathcal{R}}}(t)\rangle  &\!=&\!
	\left[\op{H}_0    
  \!-\!\sum_{\mu,\mu^\prime}\!\mathcal{D}^{(1)*}_{\mu,\mu^\prime}\!(\gamma_{\mathcal{R}})\,
	\text{E}_{\mu^\prime}(t)\,\op{r}_{\mu}\right]|\Psi^{N}_{\gamma_{\mathcal{R}}}(t)\rangle\,,\nonumber\\\vspace{-0.3cm}
\end{eqnarray}
with $\op{H}_0$ the field-free Hamiltonian, $\gamma_{\mathcal{R}}$ the Euler angles defining the orientation of the
molecular frame, $\mathcal{R}$, with respect to the laboratory frame $\mathcal{R}^\prime$, $\mathcal{D}^{(1)}_{\mu,\mu^\prime}(\gamma_{\mathcal{R}})$ the elements of the Wigner rotation matrix,  $\text{E}_{\mu}(t)$ the electric field component along the spherical unit vector ${\text{\textbf{e}}}_{\mu}$, with $\mu=\pm 1,0$, and finally, $\op{r}_{\mu}$ the component of the position operator in the direction of ${\text{\textbf{e}}}_{\mu}$.
In $\mathcal{R}^\prime$, the electric field is a superposition of an XUV comb and an IR, resp. visible, field,
\begin{eqnarray} 
\text{\textbf{E}}(t) 
	=  \sum^{N_{p}}_{k}\text{\textbf{E}}^{(\text{\tiny XUV})}_{k}(t) 
 + \text{\textbf{E}}^{(\text{\tiny{IR/VIS}})}(t-\tau_{_{\text{IR/VIS}}})\,,\quad\quad  
\end{eqnarray} 
with
\begin{eqnarray}
  \nonumber
  \text{\textbf{E}}^{(\text{\tiny XUV})}_{k}(t) =
  h_{k}(t-\tau_{k}) 
  \label{eq:xuvpulse}
  \mathrm{Re}\left\{\text{E}_{k,0}\, e^{-i(\omega_k(t-\tau_k)+\tilde{\phi}_k)}\,
 \text{\textbf{e}}_{\text{\tiny{\textsc{XUV}}}}\right\},
\end{eqnarray}
where $k=\!2q+1\!$ with $q$ an integer,
i.e., the XUV frequencies are odd multiples of the fundamental $\omega_0$ such that $\omega_{2q+1}\!=\!(2q+1)\omega_0$.
$h_{k}(\cdot)$ is a Gaussian function
    and $\text{\textbf{e}}_{\text{\tiny{\textsc{XUV}}}}$  
  the covariant spherical unit vector describing circularly left ($\mathbf{e}_{-1}=(\mathbf{e}_{x}-i\mathbf{e}_y)/\sqrt{2}$), 
  right ($\mathbf{e}_{+1}=-(\mathbf{e}_{x}+i\mathbf{e}_y)/\sqrt{2}$) and linear
    ($\mathbf{e}_{0}=\mathbf{e}_z$) 
  polarization of the radiation field.
  Note that $\tau_{k}$  describes the inherent chirp if the various XUV components are  temporally delayed~\cite{PhysRevA.69.063416,mairesse2003attosecond}. 
  IR and visible pulses are analogously defined with 
  $h_{_{\text{IR/VIS}}}(t-\tau_{_{\text{IR/VIS}}})$ and $\omega_{_{\text{IR/VIS}}}$, and
  $\tau_{_{\text{IR/VIS}}}$ describes 
  the delay between the IR/VIS and XUV field.
  
  In the following, we explore two cases. First, the frequency $\omega_{_{\text{IR}}}$ is identical
  with
  the fundamental $\omega_0$ ($\omega_{_{\text{IR}}}\!=\!\omega_0\!=\!1.5\,\mathrm{eV}$) to control PECD at the sidebands,
  cf.~Fig.~\ref{fig:figure01}(a). Second, the frequency is twice the fundamental ($\omega_{_{\text{VIS}}}=3.0\,\mathrm{eV}$) to control PECD at the harmonic peaks,
  cf.~Fig.~\ref{fig:figure03}(a).
\end{subequations}
The PECD is determined by the difference
in the photoelectron angular distribution (PAD) obtained
by reversing the helicity of the XUV, the IR, or both, 
\begin{eqnarray} 
\label{eq:mypecd}
  \text{\textsc{PECD}}(\epsilon_k, \theta_{\boldsymbol{k}^\prime}) =
    \dfrac{d^2\sigma(\textbf{e}_{\text{\tiny{\textsc{XUV}}}},\textbf{e}_{\text{\tiny{\textsc{IR}}}})}{d\epsilon_k\,d\Omega_{\boldsymbol{k}^\prime}}
   -\dfrac{d^2\sigma(\textbf{e}^\prime_{\text{\tiny{\textsc{XUV}}}},\textbf{e}^\prime_{\text{\tiny{\textsc{IR}}}})}{d\epsilon_k\,d\Omega_{\boldsymbol{k}^\prime}} \,.
\end{eqnarray}
Throughout, we report PECD in percentage of the mean peak intensity of the PADs obtained with polarization reversals~\cite{PowisJCP2000}. 
Following Refs.~\cite{GoetzPRL2019,GoetzJCP2019}, the orientation-averaged photoelectron momentum distribution 
is obtained by calculating
$|\Psi^{N}_{\gamma_{\mathcal{R}}}(t)\rangle$ within the time-dependent configuration interaction singles method 
and second-order time-dependent perturbation theory of the light-matter interaction. The differential photoionization cross section is then given by
\begin{eqnarray}
 \label{eq:SecondOrderApprox}
  \dfrac{d^2\sigma(\textbf{e}_{\text{\tiny{\textsc{XUV}}}},\textbf{e}_{\text{\tiny{\textsc{IR}}}})}{d\epsilon_k\,d\Omega_{\boldsymbol{k}^\prime}} &\approx&\!\!  \int\!
    \big|\alpha^{\boldsymbol{k}^\prime\,(1)}_{i_0}(\gamma_{\mathcal{R}})
    \!+\!\alpha^{\boldsymbol{k}^\prime\,(2)}_{i_0}(\gamma_{\mathcal{R}})  \big|^2
    \mathrm{d}^3\gamma_{\mathcal{R}},\quad\quad
\end{eqnarray} 
with $\alpha^{\boldsymbol{k}^\prime\,(1,2)}_{i_0}(\gamma_{\mathcal{R}})\!=\!\langle\Phi^{\boldsymbol{k}^\prime}_{i_0}|\Psi^{\!N(1,2)}_{\gamma_{\mathcal{R}}}\!(t)\rangle$ 
the first, resp. second, order corrections for $t\to\infty$. The index $i_0$ refers to the contribution from
the highest occupied molecular orbital (HOMO). 
  
  \begin{figure}[!tb]
  \centering
  \includegraphics[width=0.99\linewidth]{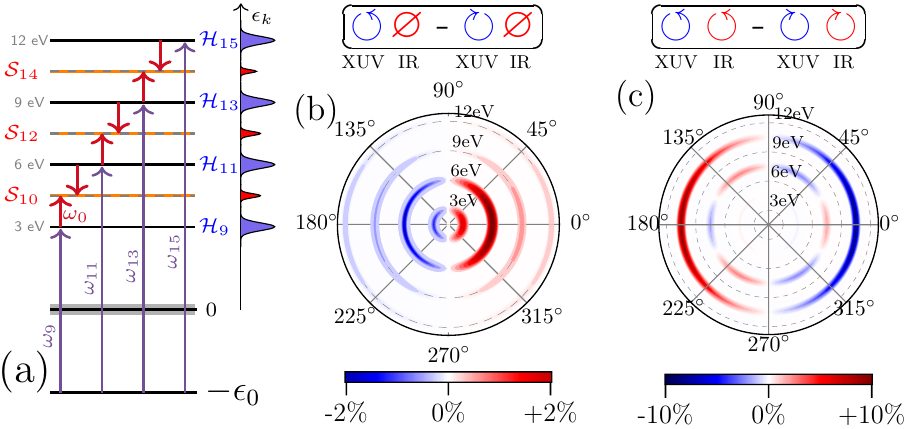}
  \caption{(a) RABBITT scheme for measuring PECD, Eq.~\eqref{eq:mypecd}:
   An asymmetric photoelectron spectrum peaked at the harmonics $\mathcal{H}_{2q+1}$ is generated by single-photon ionization of a chiral molecule's ground state (with ionization potential $\epsilon_0$), driven by an XUV comb with frequencies $\omega_{2q+1}$, which are odd multiples of $\omega_0$. In the presence of an IR field with mean photon energy $\omega_{_{\text{IR}}}\!=\!\omega_{0}$, portions of the photoelectron wave packet with energies $\mathcal{H}_{2q-1}$ and $\mathcal{H}_{2q+1}$ interfere, giving rise to sideband formation $\mathcal{S}_{2q}$. This interference can be used to modify the magnitude of PECD, seen most prominently when comparing PECD obtained with (c) and without (b) IR field.
 }
  \label{fig:figure01}
  \end{figure}
The continuum states $|\Phi_{i_0}^{\boldsymbol{k}^\prime}\rangle$, consisting of particle-hole excitations to the continuum orbitals $\varphi^-_{\boldsymbol{k}^\prime}(\boldsymbol{r})$ with energies $k^2/2$ 
and direction of photoelectron emission $\Omega_{\boldsymbol{k}^\prime}\!=\!(\theta_{\boldsymbol{k}^\prime}, \phi_{\boldsymbol{k}^\prime})$ in $\mathcal{R}^\prime$~\cite{footnote2} 
are constructed from the antisymmetrized  product~\cite{mccurdy2001practical,baertschy2001accurate} 
\begin{eqnarray}
 \Phi^{\boldsymbol{k}^\prime}_{i_0}(\boldsymbol{r}_1,\dots \boldsymbol{r}_N ) &=& 
\mathcal{A}_{N}\big[\varphi^-_{\boldsymbol{k}^\prime}(\boldsymbol{r});\Phi_{i_0}(\boldsymbol{r}_{1},\dots \boldsymbol{r}_{N-1})\big]\,,
\end{eqnarray}
with $\Phi_{i_0}(\boldsymbol{r}_{1},\dots \boldsymbol{r}_{N-1})$ the ionic component and $\varphi^{-}_{\boldsymbol{k}^\prime}(\mathbf{r})$ the
photoelectron scattering wave function. The latter is obtained as a solution of the scattering problem~\cite{LucchesePRA1982},
\begin{subequations}
\begin{eqnarray}
  \left[-\dfrac{\boldsymbol{\nabla}^2}{2}-\dfrac{1}{r}\! +V(\boldsymbol{r})-\dfrac{k^2}{2} \right]\varphi^{-}_{\boldsymbol{k}}(\boldsymbol{r})&=&0\,,
\end{eqnarray}
with $V(\boldsymbol{r})$ the short-range portion of the static-exchange potential~\cite{LucchesePRA1982}. The scattering states with momentum $\boldsymbol{k}^\prime$ in 
the laboratory frame are obtained according to 
\begin{eqnarray}
\varphi^{-}_{\boldsymbol{k}^\prime}(\boldsymbol{r}) &=& 
\!\sum_{\ell,m,m^\prime}\!\!\!\!\varphi^-_{k,\ell,m}(\boldsymbol{r})\,
\mathcal{D}^{(\ell)\dagger}_{m, m^\prime}(\gamma_{\mathcal{R}})
 Y^{\ell *}_{m^\prime}(\Omega_{\boldsymbol{k}^\prime})\,,\quad 
\end{eqnarray}
\end{subequations}
with $\varphi^{-}_{k,\ell,m}\!=\!\langle\ell,m|\varphi^{-}_{\boldsymbol{k}}(\boldsymbol{r})\rangle$.
The theoretical challenge in describing continuum-electron interferometry stems
from calculating matrix elements between continuum states. To this end,
some schemes use regularization~\cite{douguet2018PRA}. Here, we choose to converge the corresponding matrix elements with respect
to the Gaussian basis. This approach may be converged systematically and
takes advantage of the fact that the perturbed wave functions
are semi-localized due to operation by the dipole (or other interaction) operator $\hat{\mu}$,
\begin{eqnarray}
	\psi^{(k)}(t) &=& (-i)^k \int_{t> t_1\dots > t_k > t_0}\!\!\!\!\!\!\!\!\!\!\!\!\!\!\!\!\!\!\!\!\!\!\!\!\!\!\!
	e^{-i\hat{H}_0 (t-t_1)}\,\hat{\mu}\,e^{-i\hat{H}_0 (t_1-t_2)}\quad\quad\\[0.1cm]
	&&\quad\quad\quad\times\dots\hat{\mu}\, e^{-i\hat{H}_0 (t_k-t_0)}\,\psi{(t_0)}\,\, {dt_1\ldots dt_k}\,.\nonumber  
\end{eqnarray}
While each operation with $\hat{\mu}$ extends the range of the function, it is still local and representable by Gaussian functions, so long as they are converged. Calculations
of occupied and unoccupied Hartree-Fock orbitals of the equilibrium geometry were obtained using the \texttt{MOLPRO} program package~\cite{werner2012molpro,werner2012molprowires} 
using the augmented correlation consistent basis set \texttt{aug-cc-pVQZ}~\cite{KendallJCP1992}. Transition matrix elements 
to scattering states were obtained with the \texttt{ePolyScat} program package~\cite{GianturcoJCP94,NatalenseJCP99,Greenman2017variational}. 

\begin{figure}[!tb]
\centering
\includegraphics[width=0.99\linewidth]{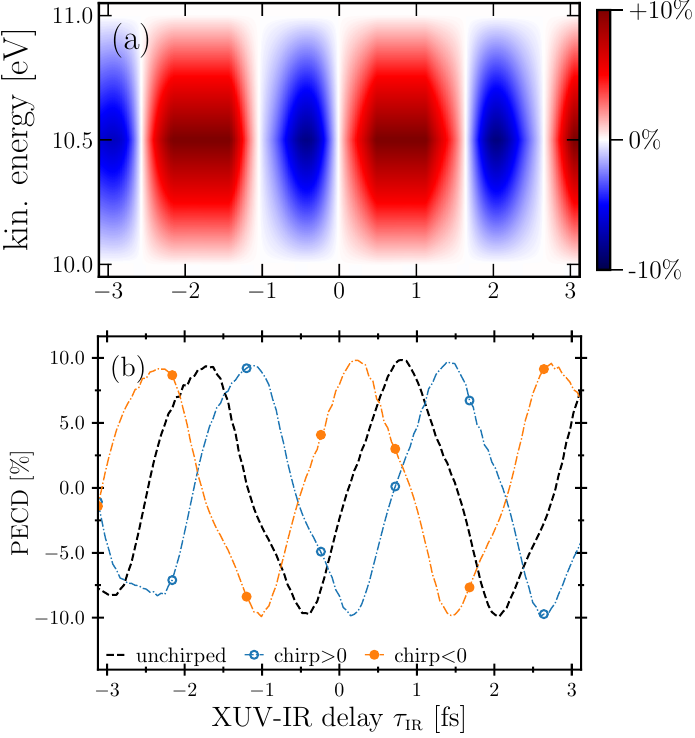}
	\caption{Robustness of RABBITT-PECD: (a) PECD trace 
         along the sideband $\mathcal{S}_{14}$ of Fig.~\ref{fig:figure01}(a) 
	 as a function of the XUV-IR time delay at the emission direction $\theta_{\boldsymbol{k}^\prime}=0^\circ$ 
         obtained with identical circular polarization states for the XUV and IR fields with polarization reversal as in  
        Fig.~\ref{fig:figure01}(c).	
	(b) Influence of the XUV comb delay dispersion on the sideband $\mathcal{S}_{14}$: 
	positive and negative chirps merely shift the PECD value 
	with respect to the unchirped case.  
	}
\label{fig:figure02}
\end{figure}

Starting with the usual RABBITT scheme depicted in Fig.~\ref{fig:figure01}(a),
we first investigate
the influence of the IR field on PECD. To this end, we optimize the IR 
pulse parameters (amplitude, duration and delay) for  
different polarization configurations of the XUV and IR pulses, with fixed
IR frequency ($\omega_{\text{IR}}=\omega_{0}$) and fixed
XUV 
parameters ($10^{11}\text{W}/\text{cm}^{2}$, $10\,$fs FWHM). The XUV pulse
is defined by a synchronous frequency comb
of odd-order harmonics $\omega_{11}, \omega_{13}$ and $\omega_{15}$ as indicated in Fig.\ref{fig:figure01}(a) with flat spectral
phase ($\tau_k\!=\!\tau_{k+1}\!=\!0$).                     
By taking the fundamental $\omega_0=1.5\,$eV,  XUV inner-shell 
and IR multiphoton contributions to the photoelectron spectra
of interest
are avoided, while simultaneously placing IR-excitation of vibrational modes far from resonance~\cite{DiemJCP76}.
In the absence of the IR  field, the  XUV pulse induces the photoelectron spectrum shown in Fig.~\ref{fig:figure01}(b), with a maximum of PECD of 2\% at a photoelectron 
kinetic energy of 6 eV.  XUV-ionization in the presence of an IR pulse with optimized intensity ($1.26\times 10^{12}\,\mathrm{W}/\mathrm{cm}^{2}$) and time-delay ($0.92$ fs), results in a stronger dichroic effect with a maximum PECD of $10\%$ at the sideband corresponding to a photoelectron kinetic energy of $10.5\,$eV, as shown in Fig.~\ref{fig:figure01}(c). 
Remarkably, the PECD obtained with the optimized IR pulse vanishes at the harmonics peaks.

The enhancement in the PECD due to the presence of the IR pulse, largest at the sideband $\mathcal{S}_{14}$ (10.5 eV), cf. Fig.~\ref{fig:figure01}(c), is obtained
with circular polarization of both, XUV and IR pulses. 
The corresponding polarization reversal is indicated 
by the red (IR) and blue (XUV) arrows in Fig.~\ref{fig:figure01}(c).
Removing the IR field and blue-shifting the XUV frequencies by 1.5eV to promote single-photon ionization at the energies corresponding 
to the sidebands (4.5, 7.5 and 10.5 eV)
results in a significantly smaller PECD, cf. the Supplemental Material~\cite{SupplementalMaterial}.
The decrease of the PECD at the $\mathcal{S}_{14}$ sideband when removing the IR field and blue-shifting the XUV
excludes kinetic energy effects and confirms an interference effect as the main driver for the enhancement. 
Keeping the polarization of the XUV pulse fixed to an anticlockwise direction
and inverting only that of the IR field results in a maximum PECD of $7\%$ 
at $7.5\,$ eV, and 5\% at 10.5 eV. Such an approach may be valuable as the polarization of
the IR fields is generally easier to manipulate. 
The strongest PECD values, however, are obtained under polarization reversal of both the XUV and IR fields. This suggests that 
contributions from bound-continuum and continuum-continuum transitions obtained with polarization reversal for both fields
are combined to enhanced the overall PECD, as we will discuss in more detail below. 

The  extent to which the interferences 
are controlled by the relative 
delay between the IR and XUV fields is investigated in Fig.~\ref{fig:figure02}(a), showing the PECD trace 
centered around the sideband $\mathcal{S}_{14}$ of Fig.~\ref{fig:figure01}(a), at $10.5\,$eV, 
for the IR- and XUV-reversed polarization configurations displayed in Fig.~\ref{fig:figure01}(c).
An oscillating behavior of the PECD as a function of the XUV-IR time delay is observed with alternating PECD sign. The pattern and oscillation period
is sensitive to the 
relative polarization configuration of both fields as well as to the sideband position and IR pulse duration, cf. Supplemental Material. In particular, 
we find that the maximum achievable PECD increases for shorter IR durations and reaches
a plateau  as a function of the ratio of IR to XUV intensities (which occurs at a few percent). In all cases, the PECD exhibits
an oscillatory behavior as function of the delay $\tau_{_{\text{IR}}}$ for fixed IR amplitude and FHWM. 

\begin{figure}[!tb]
\centering
  \includegraphics[width=0.90\linewidth]{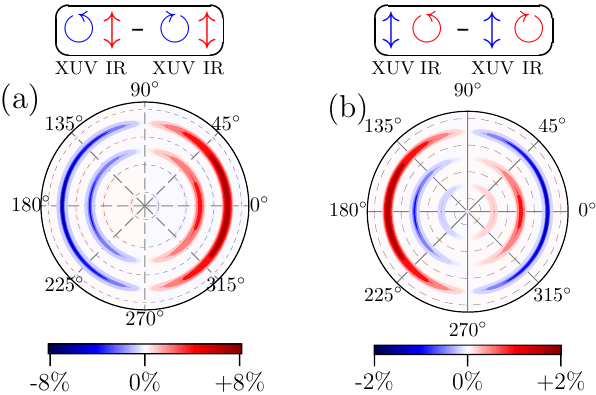}
  \caption{Disentangling bound and continuum contributions to RABBITT-PECD by choice of the polarization configurations:
  (a) PECD obtained with circularly polarized XUV and linearly polarized IR 
  pulses. (b) PECD obtained with linearly polarized XUV and circularly polarized IR  pulses. Shown is the maximum PECD obtained for the respective optimal time delays. The dotted-gray lines in the polar plots indicate the harmonic peak positions $\mathcal{H}_{2q+1}$. }
\label{fig:figure04}
\end{figure}

The PECD in Fig.~\ref{fig:figure02}(a) was obtained by assuming a flat spectral phase for the overall XUV comb, i.e., unchirped XUV harmonics.
The RABBITT technique has exposed, however,  the existence of a quadratic spectral phase for the plateau of high harmonic generation~\cite{mairesse2003attosecond,PhysRevA.69.063416,varju2005reconstruction,klunder2011probing,isinger2019accuracy}.
This has been interpreted as a temporal dispersion in the emission time of the group delay associated with each harmonic $\omega_{2q+1}$ \cite{PhysRevA.69.063416,mairesse2003attosecond}. Consequently, 
the various XUV frequency components defining the comb are linearly delayed in time~\cite{PhysRevA.69.063416}.
In Fig.~\ref{fig:figure02}(b) we investigate the extent to which such a delay dispersion affects the overall PECD signal. In detail, a positive chirp, corresponding to lower XUV harmonics preceding the higher ones,  introduces
a shift of the PECD peaks towards positive XUV-IR time-delays (blue line with empty circles) compared to the unchirped 
case (dotted-black line). Conversely, a negative chirp results in a shift towards  negative
time-delays (orange line with full circles). For a fixed delay, dispersion effects inherent to the high-order harmonic generation process may significantly affect the PECD strength. 
Remarkably,
the chirped and unchirped XUV combs generate just delayed versions of the time-resolved PECD.  
This points to the 
robustness of the proposed interferometric scheme: If the 
XUV delay dispersion merely introduces a shift in the PECD oscillations, for an 
unknown chirp, the maximum 
achievable PECD  can simply be retrieved by 
scanning the XUV-IR time-delay, cf. Fig.~\ref{fig:figure02}(b).

\begin{figure}[!tb]
\centering
\includegraphics[width=0.95\linewidth]{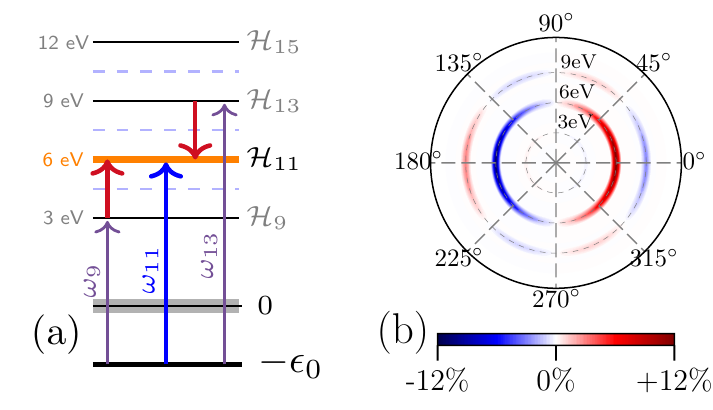}
	\caption{(a) Modified RABBITT scheme for probing the kinetic energy dependence of PECD: Replacing the IR pulse by a visible field (with
	$\omega_{\text{\textsc{VIS}}}\!=\!2\omega_0$) results in interferences between even- and odd-parity pathways and controls PECD at the harmonic peaks instead of the sidebands.
	(b) Corresponding PECD to be compared to PECD without interference at the harmonic peaks (Fig.~\ref{fig:figure01}(b)) and to interference-enhanced PECD at the sidebands (Fig.~\ref{fig:figure01}(c)). 
	}
\label{fig:figure03}
\end{figure}

Modifying the polarization configurations allows us to harness the full potential of the interferometric scheme. To this end,
we evaluate the PECD  with left/right polarization reversal for the XUV 
field, keeping  the IR polarization  linear, as indicated by the arrows in Fig.~\ref{fig:figure04}(a). In the polarization configuration of Fig.~\ref{fig:figure04}(a),
the role of the IR field is reduced to merely promoting interferences between two consecutive 
harmonic components of the photoelectron wave packet generated 
in single-photon ionization by the XUV comb. In this scenario, continuum-continuum transitions do not contribute to the PECD since the IR is linearly polarized.
As before,  the interferences can be controlled by the XUV-IR time delay, which for this configuration results in a maximum PECD of 8\%, shown in Fig.~\ref{fig:figure04}(a), at a photoelectron kinetic energy of 10.5 eV. This has to be compared to the PECD of just 2\% 
in the absence of the IR field 
at the same photoelectron energy, obtained with appropriately shifted
 XUV frequencies, cf. Supplemental Material.

The enhancement of the PECD 
highlights the role of interference
between continuum states, even for a linearly polarized IR pulse.  The difference between 10\% in Fig.~\ref{fig:figure01}(c) and 8\%,
in Fig.~\ref{fig:figure04}(a) may be attributed to chiral effects directly in the continuum. 
To 
directly probe the chirality of the continuum states, we show in
  Fig.~\ref{fig:figure04}(b) the PECD
  for fixed linear polarization of the XUV field and
 circular polarization of the IR pulse. A 
maximum PECD of 2\% is found at a photoelectron energy of 10.5 eV. 
The PECD of 2\% shown in Fig.~\ref{fig:figure04}(b) is exclusively due to the continuum states as bound states do not contribute to the PECD for linearly polarized XUV fields.

Finally, we modify the 
standard RABBITT scheme depicted in Fig.~\ref{fig:figure01}(a) to promote interferences between harmonics separated by
$\omega_{\text{VIS}}\!\!=\!\!2\omega_{\text{IR}}=3.0\,$eV as indicated in Fig.~\ref{fig:figure03}(a).  Optimizing the
visible pulse
parameters (amplitude, FWHM, time delay) while keeping those of the XUV as in Fig.~\ref{fig:figure01} and with the relative polarization configuration of Fig.~\ref{fig:figure01}(c), 
we obtain a maximum 
PECD of about 12\% at a photoelectron kinetic energy of 6 eV 
shown in Fig.~\ref{fig:figure03}(b). 
Analogously to Fig.~\ref{fig:figure02}(b), the magnitude of PECD oscillates as a function of the VIS-to-XUV time delay.
The maximum PECD of 12\% at 6 eV in Fig.~\ref{fig:figure03}(b) has to be compared to the PECD of 3\% at 6 eV obtained with the XUV alone shown in Fig.~\ref{fig:figure01}(b).
The significant enhancement in PECD compared to Fig.~\ref{fig:figure01}(b) can be attributed to the interfering pathways
$\omega_9\!+\!2\omega_{\text{IR}}$ and $\omega_{13}\!-\!2\omega_{\text{IR}}$ depicted by the violet and red arrows in Fig.~\ref{fig:figure03}(a), which are absent in Fig.\ref{fig:figure01}(b). In fact, the single-photon ionization pathway represented by $\omega_{11}$ in Fig.~\ref{fig:figure03}(a) plays a minor role in the PECD enhancement: removing $\omega_{11}$ from the XUV comb
 results in maximum PECD of
10\%. The main contribution to the PECD enhancement observed in Fig.~\ref{fig:figure03} can thus be attributed to a two-photon pathway interference effect, determined by the pathways
$\omega_9\!+\!2\omega_{\text{IR}}$ and $\omega_{13}\!-\!2\omega_{\text{IR}}$ interfering at 6 eV.
Replacing the IR field ($\omega_0=1.5$ eV) that maximizes the PECD at the sidebands in Fig.~\ref{fig:figure01}(c) with a visible field ($\omega_0=3.0$ eV) as in Fig.~\ref{fig:figure03} allows us to probe and exploit kinetic energy dependence effects. Both two-photon interferometric schemes discussed here result in an overall enhancement of the PECD compared to the XUV-only case. This attests to the ability of two-photon quantum pathway interferences to enhance PECD across the kinetic energy spectrum by exploiting continuum-continuum transitions with a minimum number of control parameters, and offers a complementary control protocol to schemes based on bound-state one- and two-photon quantum interference.

In conclusion, using $\ce{CHBrClF}$ as a chiral  prototype, we have identified a robust scheme  
for the control of photoelectron circular dichroism in randomly oriented chiral molecules
based on laser-assisted photoelectron wave packet interferometry.
Control of PECD is achieved by manipulating interferences which can be controlled by means of the XUV-IR time delay. 
In striking contrast to multicolor phase control schemes~\cite{GoetzPRL2019} which are highly sensitive to the temporal coherence and require
pulse shaping techniques which, while within experimental reach, are not yet practically accessible,
the present control scheme considerably reduces the number of control parameters while simultaneously
preserving the robustness with respect to the overall spectral phase. In this regard, we have analyzed the 
sensitivity of the PECD with respect to the inherent XUV group delay dispersion,
and shown that the XUV comb delay dispersion simply introduces a shift in the PECD oscillations 
as a function of the XUV-IR time delay but does not affect the maximum PECD amplitude. It is thus sufficient to scan the XUV-IR time delay
to retrieve the optimal delay for a specific dispersion of the consecutive odd-order harmonics contributing to the same sideband energy.
We have also shown that by controlling the relative polarization between the XUV and IR pulses, the RABBITT technique can be exploited to probe
the individual contributions of bound and continuum states to the PECD in a time-resolved manner.
In this regard, experimental verification of RABBITT-PECD would confirm that PECD is largely due to 
the chiral potential experienced by the bound state electrons while continuum-state transitions contribute to a lesser extent. The continuum states, however, are critical to enhance  PECD via interferences promoted by the IR field.
Our work opens the way for robust control of PECD based on 
widely implemented laser-assisted photoelectron wave packet interferometry techniques with a minimum number of control parameters while disentangling the bound and continuum contributions
to the PECD.

\begin{acknowledgments}
We acknowledge Chi-Hong~Yuen for his contribution to implementing part of the code used in this work. 
The computing for this project was performed on the Beocat Research Cluster at 
Kansas State University, which is funded in part by NSF grants CNS-1006860, EPS-1006860, and EPS-0919443, ACI-1440548, CHE-1726332, and NIH P20GM113109,
and used resources of the National Energy Research Scientific
Computing Center (NERSC), a U.S. Department of Energy Office of Science User
Facility operated under Contract No. DE-AC02-05CH11231 using NERSC award BES-ERCAP0024357. C. A. and L. G. were supported by the Chemical Sciences, Geosciences, and Biosciences Division, Office of Basic Energy Sciences, Office of Science, U.S. Department of Energy, under Grant No. DE-SC0022105. C. A. also acknowledges NSF Grant No. 2244539 for additional support during the summer of 2023. A. B. and C. P. K. acknowledge financial support from the Deutsche Forschungsgemeinschaft (CRC
    1319). 
\end{acknowledgments}

\section{\textsc{Supplemental material}}
We provide here: 
(i) convergence test calculations of the RABBITT-PECD 
as a function of the basis set used for the electronic structure 
calculations,
(ii) calculations of PECD as a function of the full width at half maximum (FWHM) of the IR field, and 
(iii) the input file parameters for the electronic structure and scattering
calculations. 
In accordance  
with the manuscript, the PECD values are given in percentage of the mean peak 
intensity of the photoelectron angular distribution (PAD).
In accordance  
with the manuscript, the PECD values are given in percentage of the mean peak 
intensity of the photoelectron angular distribution (PAD).

\subsection{\textsc{Convergence tests}}
\subsection{One-photon PECD}
\label{subsec:1ph-pecd}

The results reported in the manuscript were obtained with the augmented correlation 
consistent basis sets \texttt{aug-cc-pVQZ}~\cite{augccpvdz} to describe occupied and unoccupied 
Hartree-Fock orbitals.                                                           
Figure~\ref{fig:figure01} displays the one-photon PECD
as a function of the photoelectron kinetic energy 
obtained with the augmented correlation consistent basis \texttt{aug-cc-pVTZ} (circled blue-dashed lines)
and the \texttt{aug-cc-pVQZ} basis (black-dashed lines). The ionizing field is defined by an isolated XUV comb   
of mean photon energy $\omega_{\text{XUV}}=\epsilon_k+|\epsilon_0|$, with $\epsilon_k$
the photoelectron kinetic energy and $\epsilon_0$ the \textsc{HOMO} energy   
obtained for each of these basis sets, 
with values of $-11.7062\,$eV and $-11.7007\,$eV at the equilibrium geometry for the  
\texttt{aug-cc-pVTZ} and \texttt{aug-cc-pVQZ} basis, respectively.
The remaining XUV pulse parameters are fixed
and correspond to those described in the manuscript. For both basis sets,
the PECD is obtained with the polarization reversal corresponding to Fig.~1(b) in the
manuscript. The IR field is absent. Overall,  
good agreement is obtained using both basis sets over the photoelectron energy range 
we have considered, which ranges from $0.5\,$eV to $12.5\,$eV. The asymptotic momentum 
components of the scattering states are truncated at 
$\text{\texttt{LmaxK}}$. For both basis sets, the PECD is found to converge
within a precision below $0.1\%$ for $\text{\texttt{LmaxK}}=20$ for all photoelectron kinetic energies we have considered.
\begin{figure}[!tb]
\centering
\includegraphics[width=0.81\linewidth]{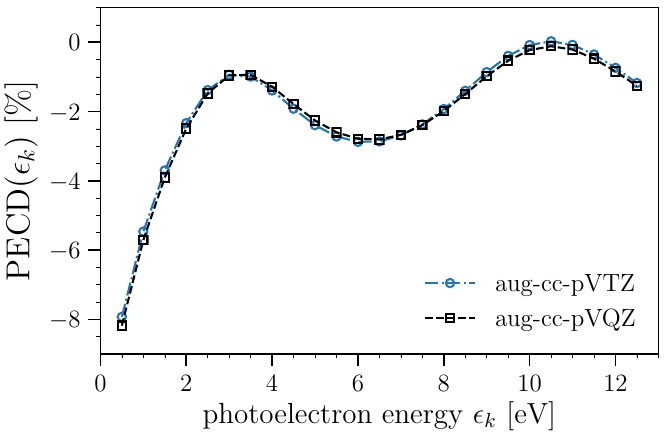}
\caption{One-photon PECD as a function of the photoelectron kinetic energy
 $\epsilon_k=\omega_{\text{XUV}}-|\epsilon_0|$ obtained with the \texttt{aug-cc-pVTZ} 
 and \texttt{aug-cc-pVQZ} basis set.  
 }
\label{fig:figure01}
\end{figure}

\subsection{RABBITT-PECD}
\label{subsec:rabbit-pecd}

In our calculations, the final continuum states populated after absorption of an XUV and IR photon
are obtained using the scattering formalism described in the manuscript. In contrast, 
the transient intermediate continuum states populated after XUV ionization are described by means of the highly-lying unoccupied Hartree-Fock orbitals
obtained from electronic structure calculations (see discussion in the manuscript). 

To check the sensitivity of the RABBITT-PECD
on the high-lying orbitals representing the intermediate continuum states, we 
compare in Fig.~\ref{fig:figure02} the RABBITT-PECD
as a function of the XUV-IR time delay obtained with the augmented  
basis sets \texttt{aug-cc-pVQZ} and \texttt{aug-cc-pVTZ}.
Black-dashed lines in Fig.~\ref{fig:figure02} correspond to the RABBITT-PECD 
as a function of the XUV-IR time delay reported in Fig.~2(b) in the manuscript for the unchirped case
(obtained with the 
\texttt{aug-cc-pVQZ} basis set). The blue-dashed lines in Fig.~\ref{fig:figure02}  
corresponds to the time-resolved RABBIT-PECD obtained with the less accurate \texttt{aug-cc-pVTZ} basis set.
While for the one-photon PECD in Fig.~\ref{fig:figure01} both basis sets are in very good agreement, 
the two-photon PECD in Fig.~\ref{fig:figure02} shows discrepancies between both basis sets.

\begin{figure}[!tb]
\centering
\includegraphics[width=0.81\linewidth]{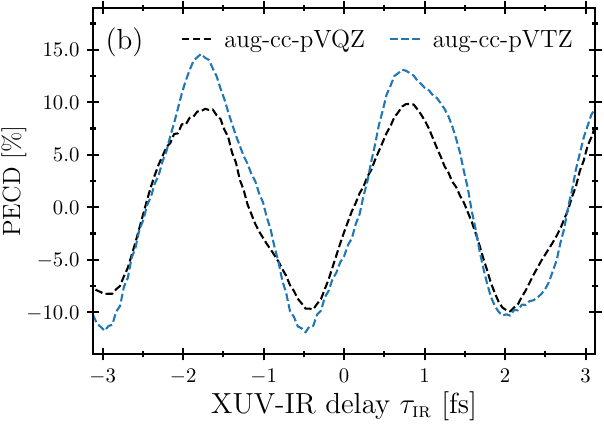}
\caption{Two-photon PECD as a function of XUV-IR delay obtained with the aug-cc-pVTZ and aug-cc-pVQZ basis sets. 
 The PECD is given in percentage of the mean peak intensity of the PADs 
 obtained with the polarization reversal configuration of Fig.~1(b) in the 
 manuscript. }
\label{fig:figure02}
\end{figure}
\label{subsec:1ph-pecd}

One source of discrepancy stems from the fact that the \texttt{aug-cc-pVQZ} basis leads to a more energetically dense set of virtual Hartree-Fock 
orbitals. 
For \texttt{aug-cc-pVQZ}, the energy distribution increases quadratically for the lower virtual Hartree-Fock orbitals 
as a function of the orbital energy (analogously to 
the energy distribution of a particle in a box) 
compared to the \texttt{aug-cc-pVTZ} basis, which has a less dense and more oscillatory energy distribution.
In either case, the virtual orbitals contributes coherently to the PECD within the bandwidth of the XUV and IR pulses
and any difference in the description of these virtual states will affect the PECD at the sidebands. 
Regardless of these small differences, the oscillations
as a function of the IR-XUV delay as well as the robustness with respect to XUV delay dispersion (chirp) persist independently of the choice of the basis.

\begin{figure}[!tb]
\centering
\includegraphics[width=0.83\linewidth]{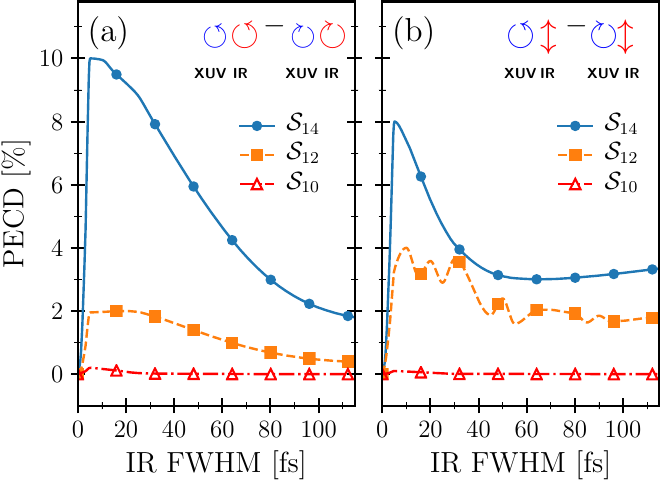}
\caption{ Maximum PECD at sidebands $\mathcal{S}_{10}$ (4.5 eV), $\mathcal{S}_{12}$ (7.5 eV) and $\mathcal{S}_{14}$ (10.5 eV) as a function of the IR full-width at half maximum (FWHM) 
for the polarization configuration indicated by the blue (XUV) and red (IR) arrows in panels (a) and (b).}
\label{fig:figure03}
\end{figure}

\subsection{\textsc{Influence of the FWHM}}

The oscillations observed in Fig.~2
in the manuscript deviate from the standard RABBITT oscillations 
which follow a pure oscillatory behavior
$\propto\cos(2\omega_{\text{IR}}\tau_0+\phi)$. The latter expression is obtained under the
assumption that the IR field can be described by a cw field. Indeed, we find that constraining
the FWHM of the IR pulse during optimization to larger values results
in a delay-dependent RABBITT-PECD described by the expected oscillatory behavior.
However, the maximum PECD strength that can be obtained with longer IR pulses 
is weaker compared to the PECD obtained with shorter IR pulses. This can be seen
in Fig.~\ref{fig:figure03}, which shows the maximum PECD achievable 
as a function of the FWHM for the XUV and IR polarization configurations indicated by the blue (XUV) and red (IR) arrows in panels (a) and (b). 
When the PECD is obtained upon polarization reversal of both XUV and IR pulses,
the resulting PECD decreases monotonically as a function of the IR FWHM from its maximum value of $10\%$. On the other hand, when the IR is linearly polarized,
the PECD decreases and reaches a plateau for the sidebands  $\mathcal{S}_{12}$ (7.5\,eV) and $\mathcal{S}_{14}$ (10.5\,eV). The difference in magnitude between the PECD change over the consider FWHM values between both pulses counter-rotating in Fig.~\ref{fig:figure03}(a) and linearly polarized IR in Fig.~\ref{fig:figure03}(b) may be understood by considering that the XUV pulse contributes most of the PECD effect as discussed in the manuscript. The relative change in magnitude is not large, but the rest of the small effect can be attributed to the polarization of the IR.
\begin{figure}[!ht]
\centering
\includegraphics[width=0.87\linewidth]{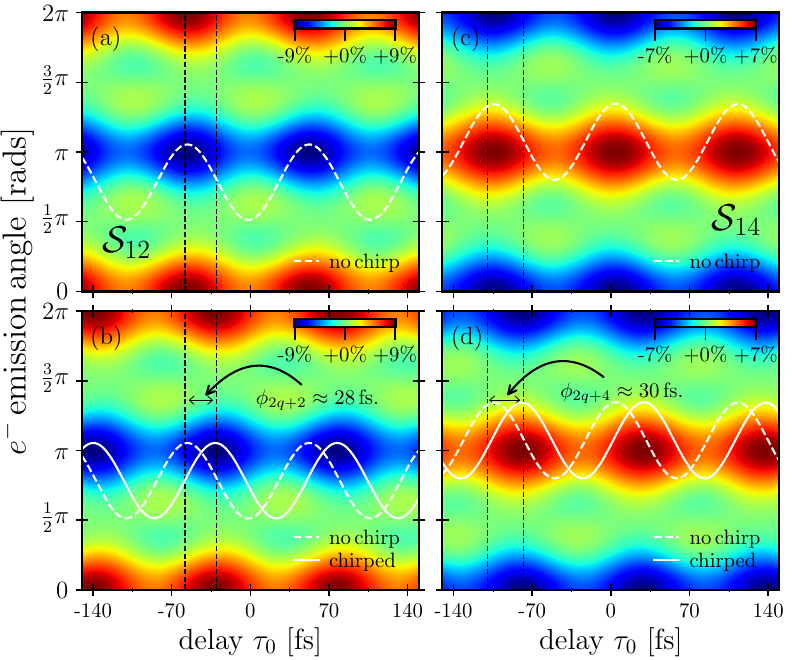}
\caption{ (a) Maximum PECD at sideband $\mathcal{S}_{12}$ as a function of the azimuthal emission angle $\theta_{\boldsymbol{k}^\prime}$ and time delay $\tau_0$ for zero group dely 
dispersion ($\tau_{j+1}-\tau_j = 0$) of the XUV field. 
(b) same as (a) but including a positive chirp in the group delay dispersion of the XUV field comb as discussed in the manuscript. The spectral 
properties of the XUV components introduces a shift in the PECD value that can be retrived by shifting the XUV-IR time delay $\tau_0$. 
Panels (c) and (d) shows the same property, for sideband $\mathcal{S}_{14}$. Note that the phase shift $\phi_{2(q+n)}$ relative to the unchirped case
is sideband-dependent.}
\label{fig:figure04}
\end{figure}

Thus, apart from introducing deviations to the standard RABBITT oscillatory behavior, shorter IR pulses lead to a stronger dichroic effect.
Pure oscillatory behavior $\propto\cos(2\omega_{\text{IR}}\tau_0+\phi)$  are restored for longer IR FWHM as shown in Fig.~\ref{fig:figure04},
showing the RABBITT oscillations (dotted-white lines) as a function of the photoemission angle (vertical axis) and the XUV-IR time delay (horizontal axis)
at fixed photoelecton energies of 7.5 eV (a) and 10.5 eV (c) for the unchirped case, respectively. Panels (c) and (d) show the PECD trace obtained with
a chirped XUV comb of odd-order harmonics. As for the shorter IR cases discussed in the manuscript, a group delay dispersion in the ionizing XUV field
merely shifts the PECD values independently of IR pulse duration as shown in panels (b) and (d) in Fig.~\ref{fig:figure04}. This highlights the 
robustness of the proposed interferometric approach as the maximum PECD strength depends on the pulse duration and the XUV-to-IR time delay and avoids
pulse shaping and control over the spectral properties of the PECD-inducing field.

\subsection{Input file parameters}

\subsection{Input file parameters for \texttt{MOLPRO} }
\label{sec:section4}
\texttt{\hspace{-0.3cm}$\#$ MOLPRO Program package~\cite{werner2012molpro,werner2012molprowires} input file}
\medskip

\noindent\texttt{***,CHFClBr gs}\\
\texttt{RCH=                 1.088 ANG}\\
\texttt{RCCl=                1.745 ANG}\\
\texttt{RCBr=                1.928 ANG}\\
\texttt{RCF=                 1.356 ANG}\\
\texttt{aHCF=              108.806 DEGREES}\\
\texttt{aHCCl=             108.5   DEGREES}\\
\texttt{aHCBr=             108.5   DEGREES}\\
\texttt{aFCCl=             109.93  DEGREES}\\
\texttt{aFCBr=             108.95  DEGREES}\\
\texttt{aClCBr=            112.09  DEGREES}\\
\noindent\texttt{geometry=\{}\\
\texttt{ C1;}\\
\texttt{ H1,C1,RCH;}\\
\texttt{ Cl2,C1,RCCl,H1,aHCCl;\}}\\
\texttt{ Br3,C1,RCBr,H1,aHCBr,Cl2,aClCBr,1;}\\
\texttt{ F4,C1,RCF,H1,aHCF,Cl2,aFCCl,-1;}\\
\texttt{\}}\\
\texttt{ basis=avqz}\\
\texttt{ \{rhf; wf,68,1,0;orbprint,40;\}}\\
\texttt{optg;}\\
\texttt{ put, molden, CHFClBr-avqz.molden;}\\

\subsection{Input file parameters for\,\,\,\texttt{\lowercase{e}P\lowercase{oly}S\lowercase{cat}}}
\label{sec:section5}
\texttt{\hspace{-0.3cm}$\#$ ePolyScat Program package~\cite{GianturcoJCP94,NatalenseJCP99} input file}\\
\texttt{\hspace{-0.3cm}$\#$ input file CHFClBr photoionization}\\
\texttt{$\#$\, and matrix elements calculation}\\
\medskip

\noindent\texttt{\#***,CHFClBr}\\
\texttt{LMax  30}\\                
\texttt{EMax  20}\\              
\texttt{EngForm }\\               
\texttt{ 0 3    }\\               
\texttt{FegeEng 13.0}\\            
\texttt{ScatEng .50}\\           
\texttt{LMaxK 20 }\\  
\texttt{NoSym}\\ 
\texttt{InitSym \char13 A\char13}\\              
\texttt{InitSpinDeg 1}\\            
\texttt{OrbOccInit 2 2 2 2 2 2 2 2 2 2 2 2 2 2 2 2 2 2 2 2 2 2 2 2 2 2 2 2 2 2 2 2 2 2}\\ 
\texttt{OrbOcc  2 2 2 2 2 2 2 2 2 2 2 2 2 2 2 2 2 2 2 2 2 2 2 2 2 2 2 2 2 2 2 2 2 1}\\ 
\texttt{SpinDeg 1 }\\        
\texttt{TargSym \char13 A\char13 }\\      
\texttt{ScatContSym \char13 A\char13}\\ 
\texttt{ScatSym \char13 A\char13}\\ 
\texttt{TargSpinDeg 2 }\\    
\texttt{IPot 10.98}\\        
\medskip

\noindent\texttt{Convert  \char13 chfclbr-avqz.molden\char13 \char13 molden2012\char13}\\
\texttt{FileName \char13 MatrixElements\char13 \char13 MatEle.idy\char13 \char13 REWIND\char13}\\ 
\texttt{FileName \char13 DumpOut\char13 \char13 DumpOut.idy\char13 \char13 REWIND\char13}\\ 
\texttt{GetBlms}\\ 
\texttt{ExpOrb}\\ 
\texttt{GenFormPhIon}\\ 
\texttt{GetPot}\\ 
\texttt{DipoleOp}\\ 
\texttt{PhIon .50}\\ 
\texttt{DumpIdy \char13 MatEle.idy\char13 .50}\\


%
\end{document}